\documentclass[amssymb,twocolumn,showpacs,superscriptaddress]{revtex4-1}
\usepackage{graphicx}
\usepackage{amsmath,amssymb,amsbsy,times,hyperref}
\usepackage{epstopdf}
\usepackage[utf8]{inputenc}  
\usepackage[T1]{fontenc}

\begin{document}

\title{All silicon Josephson junctions}
\author{F. Chiodi}
\email{francesca.chiodi@u-psud.fr}
\affiliation{Centre de Nanosciences et de Nanotechnologies, CNRS, Univ. Paris-Sud, Université Paris-Saclay, C2N – Orsay, 91405 Orsay cedex, France}
\author{J.-E. Duvauchelle}
\affiliation{Univ. Grenoble Alpes, CEA-INAC-PHELIQS, F-38000 Grenoble, France}
\author{C. Marcenat}
\affiliation{Univ. Grenoble Alpes, CEA-INAC-PHELIQS, F-38000 Grenoble, France}
\affiliation{Univ. Grenoble Alpes, Institut Néel, F-38000 Grenoble, France}
\author{D. Débarre}
\affiliation{Centre de Nanosciences et de Nanotechnologies, CNRS, Univ. Paris-Sud, Université Paris-Saclay, C2N – Orsay, 91405 Orsay cedex, France}
\author{F. Lefloch}
\affiliation{Univ. Grenoble Alpes, CEA-INAC-PHELIQS, F-38000 Grenoble, France}

\date{\today}
\begin{abstract}
We have realised laser-doped all-silicon superconducting (S)/ normal metal (N) bilayers of tunable thickness and dopant concentration. We observed a strong reduction of the bilayers critical temperature when increasing the normal metal thickness, a signature of the highly transparent S/N interface associated to the epitaxial sharp laser doping profile. We extracted the interface resistance by fitting with the linearised Usadel equations, demonstrating a reduction of one order of magnitude from previous superconductor/doped Si interfaces. In this well controlled crystalline system we exploited the low resistance S/N interfaces to elaborate all-silicon lateral SNS Josephson junctions with long range proximity effect. Their dc transport properties, such as the critical and retrapping currents, could be well understood in the diffusive regime.  Furthermore, this work lead to the estimation of important parameters in ultra-doped superconducting Si, such as the Fermi velocity, the coherence length, or the electron-phonon coupling constant, fundamental to conceive an all-silicon superconducting electronics.

\end{abstract}
\pacs{74.45.+c, 73.23.-b, 73.40.Sx}
 \maketitle

\emph{Introduction --}
Quantum information is one of the main challenges in the search of faster, more efficient, data treatment. In this frame, spin qubits made of isotopically purified silicon have recently emerged as one of the most promising technologies \cite{Saeedi, Muhonen}. Indeed, silicon is an extremely well known material, where chemical purification, crystal growth, and defect control have been developed to a high extent. Similarly, micro and nano fabrication clean room processes have been optimised for decades, allowing the realisation of scalable devices. Finally, thanks to a low spin-orbit coupling and to the presence of stable isotopes without nuclear magnetic moment, it is possible to obtain inimitable electron and nuclear spin coherence times \cite{Saeedi, Muhonen}. Silicon thus appears as a choice material to realise coherent quantum circuits, even more so when coupled to superconductivity, whose robustness and absence of dissipation were proven to be a great asset in the realisation of the most advanced structures for quantum information \cite{Brecht, Devoret}.
Superconducting silicon offers a perfect solution to combine on the same chip the advantages of Si and superconductivity \cite{Tahan}. Obtained by laser doping, superconducting silicon layers can be epitaxied over a silicon substrate, form crystalline superconductor/doped Si interfaces without Schottky barriers (thanks to the comparable density of states), and ensure ohmic contacts with metals \cite{ChiodiApp}. First studies have explored the superconductivity in Si:B and detailed its dependence on the doping parameters \cite{ChiodiApp,Thierry}, and recently superconducting silicon SS'S Dayem Superconducting Quantum Interference Devices (SQUIDs) \cite{SQUID} have been realised with standard clean room processes. In this paper we demonstrate Superconductor (S)/ Normal metal (N)/ Superconductor (S) Josephson junctions, fabricated entirely from silicon laser doped SN bilayers. The bilayers' and junctions' behaviour could be modelled in the limit of diffusive superconductivity. The well controlled technology, together with the demonstration of transparent interfaces between superconductor and silicon layers, constitute an important step towards an all-silicon superconducting electronics.\\

\begin{figure}[t!]
\includegraphics[width=\columnwidth]{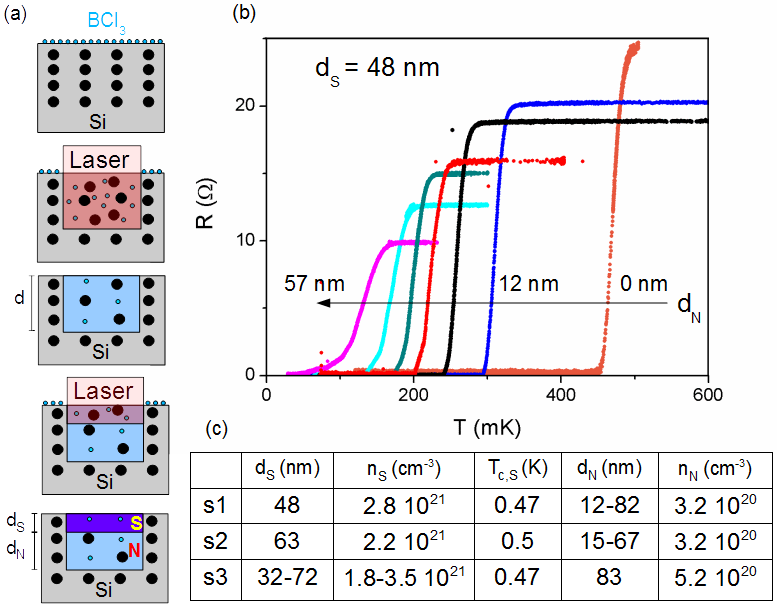}
\caption{(a) Bilayer fabrication: chemisorbtion of the precursor gas; laser melting of the substrate and dopant diffusion in the liquid phase; solidification and epitaxy of a normal metal Si:B crystal over the Si substrate; chemisorbtion and laser melting of a thin Si:B layer; epitaxy of a superconducting Si:B layer over the normal metal Si:B layer. (b) Resistive transitions $R(T)$ of SN bilayers (s1) with $d_N=0,12,17,26,32,42,57\,nm$. (c) Characteristics of the three bilayers series realised: thickness $d_S$, dopant concentration $n_S$, and critical temperature $T_{c,s}$ of the superconducting layer, thickness $d_N$ and dopant concentration $n_N$ of the normal layer. }
\label{GILD}
\end{figure}

\emph{Laser doping and sample fabrication--} The adopted doping technique, Gas Immersion Laser Doping (GILD) allows the realisation of box-like homogeneous boron doped layers of varied active concentration ($6\times 10^{18}$cm$^{-3}-6\times 10^{21}$cm$^{-3}$) and thickness (5-300 nm), with sharp, epitaxial Si/Si:B interfaces \cite{Bhaduri, Cammilleri,Marcenat}. The doping takes place in a UHV chamber ($10^{-9}\,$mbar), where a puff of the precursor gas, BCl$_3$, is injected onto the Si sample surface saturating the chemisorption sites \cite{Boulmer}. The substrate is then melted by a pulsed excimer XeCl laser of pulse duration 25 ns. The boron diffuses into the liquid silicon and is incorporated substitutionally: a Si:B crystal is epitaxied on the underlying silicon (Fig. \ref{GILD}). Because of the short pulse duration and the high recrystallization speed, boron concentrations larger than the solubility limit ($\sim$ 1$\times 10^{20}$cm$^{-3}$ \cite{Harame}) and as high as 11 at.$\%$ ($6 \times 10^{21}$ cm$^{-3}$) can be obtained without the formation of B aggregates \cite{HoummadaAPT}. 
The good control over the structural properties is reflected in the electronic properties, as homogeneous superconducting layers can be realised with superconducting critical temperatures up to 0.7 K \cite{Thierry,DahlemSTM, PLIE}. GILD thus allows us to fabricate thin layers with nanometric thickness control, a dopant concentration spanning a wide range from the semiconducting to the metallic and superconducting limits, and a finely tuned superconducting $T_c$ solely controlled by the boron dose.  \\

\emph{SN bilayers --} Laser doped superconducting (S)/normal metal (N) bilayers were fabricated from n-type Si wafers of resistivity 45 $\Omega \,$cm. The SN bilayers consist in the superposition of a superconducting layer on top of a normal (non superconducting) layer. We started the fabrication by doping a normal layer of thickness $d$ and dopant concentration $n_N$. We then reduced the laser energy, and doped a thinner layer above the superconducting threshold (Fig. \ref{GILD}-a). This results in an upper superconducting layer of thickness $d_S$ with a critical temperature $T_{c,S}$, on top of a bottom normal layer of thickness $d_N=d-d_S$ with a fixed dopant concentration $n_N$. Ti (15 nm)/Au (200 nm) contacts were then evaporated after a surface BHF deoxidization.
We realised two series of samples with constant $d_S$ and varying $d_N$, s1 and s2, and a third serie of samples with constant $d_N$ and varying $d_S$, s3 (see Tab. \ref{GILD}-c for details). \\
In all the series, the S and N layers thickness were chosen to be smaller than, or comparable to, the superconductor coherence length $\xi \sim 80\,$nm \cite{Audrey}: $d_S=32-72\,nm \lesssim \xi$ and $d_N = 12-83\,nm \lesssim \xi $. \\

\begin{figure}[t!]
\includegraphics[height=7cm, width=\columnwidth]{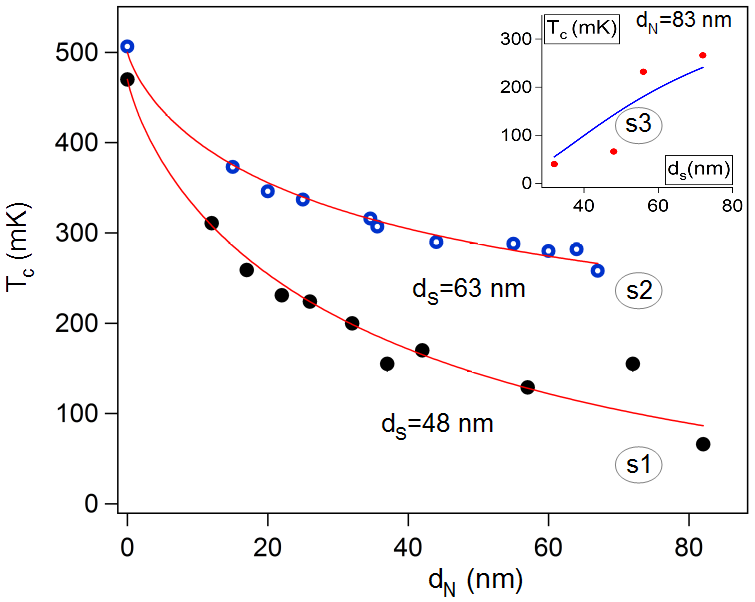}
\caption{Critical temperature $T_c$ as a function of the normal layer thickness $d_N$ for series s1 (dots) and s2 (circles). The lines are fits from eq.\eqref{usadelTc} with $T_{c,S}=0.47\,K$ and $b=0.486$ for s1, $T_{c,S}=0.5\,K$ and $b=0.52$ for s2. (Inset) Critical temperature $T_c$ as a function of the superconducting layer thickness $d_S$ for series s3 (dots), fitted with eq.\eqref{usadelTc} (line). This fit is less accurate than that of $T_c (d_N)$ due to the fact that we adopted an average $b=0.6$ for s3 series. The $b$ value actually ranges from 0.53 to 0.66, as the dopant concentration is varied in s3 in accordance to $d_S$ to insure a constant $T_{c,S}$. }
\label{Tcdn}
\end{figure}

\emph{Large $T_c$ modulation --} Characteristic resistive transitions of the s1 series SN bilayers are shown in Fig. \ref{GILD}-b. The transitions are quite sharp, their width varying between 25 and 58 mK. We observe a drastic reduction of the bilayer $T_c$ when the normal layer thickness increases. We detail the dependence of $T_c$ vs $d_N$ in Fig. \ref{Tcdn}, where we show two series of bilayers, s1 and s2, of different $d_S=48$ and $63\,$nm. We observe a monotonic $T_c$ decrease with $d_N$, followed by a saturation at large $d_N$. 
The $T_c$ reduction, as high as 86$\%$, is more marked in the series of samples with the thinner superconducting layer, where the Cooper pairs spend comparatively more time in the normal metal. 
We understand qualitatively the saturation as the Cooper pairs explore the normal layer on an average depth $\xi$: when $d_N>\xi$  the superconductivity is no longer sensitive to an increased thickness of the normal layer. \\
\emph{High interface quality --} The strong influence of the normal layer over the superconducting one supports the expectation of a transparent S/N interface. Indeed, the $T_c$ suppression with $d_N$ is determined by the interface resistance between the superconducting and the metallic doped Si layer. We could thus demonstrate an unprecedented low SN contact resistance in silicon laser doped bilayers, thus enabling long range proximity effect.

We modelled the variation of $T_c$ with $d_N$ (at fixed $d_S$) and with $d_S$ (at fixed $d_N$) with the analytical solution of the linearised Usadel equations near $T_c$ in the limit of good transparency ($\tau >> k_B T_c$) \cite{Sophie,Feigelman}:
\begin{equation}
T_c = T_{c,S} \, \left[\frac{T_{c,S}}{1.14 \,\Theta_D} \sqrt{1+\left(\frac{k_B \Theta_D}{\tau}\right)^2}\right]^{b \frac{d_N}{d_S}}
\label{usadelTc}
\end{equation}
\begin{equation}
\tau=\frac{\hbar}{2 \pi} \,\frac{v_{F,S}}{\rho_{int}}\, \frac{b \,d_N+d_S}{b \,d_N\, d_S} \qquad b=\frac{v_{F,N}}{v_{F,S}}=(\frac{n_{N}}{n_{S}})^{1/3}
\end{equation}
As most of the parameters involved in the model are well known either from direct measurements or from the literature, the interface resistance can be estimated through the fitting parameter $\rho_{int}$, the dimensionless interface resistance per channel. 
The total interface resistance per unit area is then given by taking all the conduction channels into account: $R_{int} A= h/2e^2\, (\lambda_{F,S}/2)^2 \, \rho_{int}$. The other parameters in the model are:\\
- $T_{c,S}$: critical temperature of the single superconducting layer, measured independently.\\
- $\Theta_D$: phonon energy scale in temperature units, estimated from literature at 450 K \cite{Boeri}. \\
- $b=v_{F,N}/v_{F,S}$: ratio of the Fermi velocities in the normal and superconducting layers. In the free electron model, $v_{F} \propto {k_F}=(3 \pi^2 n_B)^{1/3}$, giving $b=(n_{N}/n_{S})^{1/3}$. The carrier densities were independently determined with Hall measurements (Table in Fig. \ref{GILD}).\\
- $v_{F,S}$: Fermi velocity in the superconductor. Estimated to $v_{F,S}\sim 2.5\pm 1\times 10^{5}$m/s from independently measured Si:B parameters ($T_c=470\,$mK, resistivity $\rho\sim 110\,\mu\Omega$cm, concentration $n_S\sim 2.8\times 10^{21}\,$cm$^{-3}$ and critical field $B_{c,2}\sim 940\,$G) \cite{ShimIEEE} .\\
It is worth mentioning that the concentration at the bottom of the doped layer decreases to zero over $d' \sim$ few nanometers, that will therefore not be superconducting. The normal layer thickness $d_N$ is thus slightly larger, and $d_S$ slightly smaller, than the doped depths. The best fits were obtained with $d'=8\,$ nm for all series, in agreement with \cite{Thierry,Bhaduri}.\\
The initial hypothesis that $\tau >> k_B T_c$, is reasonably confirmed, with $\tau/k_B T_c = 3-27$ in the range studied. \\
We show in Fig. \ref{Tcdn} the good agreement obtained with the experimental data when adjusting the parameter $\rho_{int}$. 
We find $\rho_{int}=4.4$ and $R_{int} A= 8\times 10^{-10}\,\Omega\,$cm$^2$ for s1 and $\rho_{int}=10$ and $R_{int} A= 2\times 10^{-9}\,\Omega\,$cm$^2$ for s2. The interface obtained is thus extremely transparent, with a resistance smaller by an order of magnitude when compared to experimental \cite{Buonomo,Lefloch, Ruby} or theoretical \cite{Sze} determinations of silicon/metallic superconductor contacts resistance.

\begin{figure}[t!]
\includegraphics[width=\columnwidth]{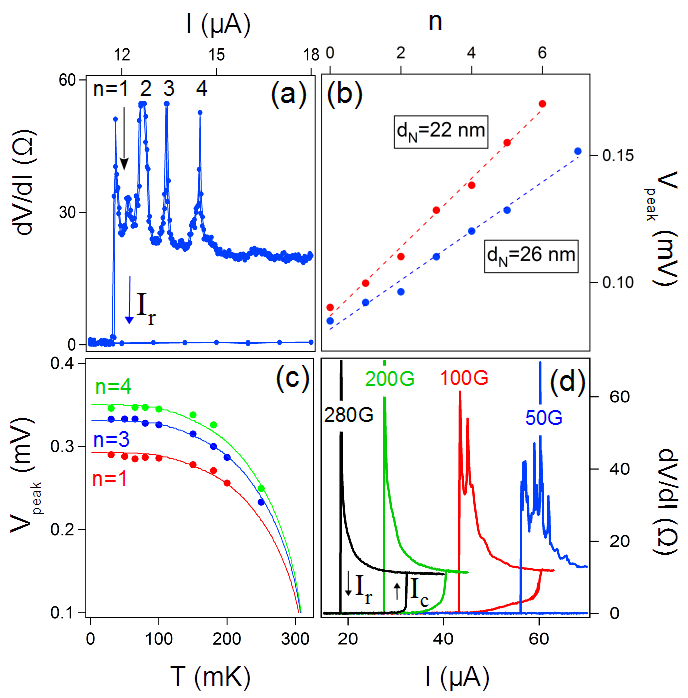}
\caption{(a) Two differential resistance curves for a bilayer with $d_S=48\,$nm and $d_N=22\,$nm at $T$=75 mK. Reproducible peaks are observed. (b) Peaks voltage in two bilayers of equal $d_S=48\,$nm and different $d_N$ at $T$=75 mK. We have arbitrarily attributed to the lower voltage peak observed the index n=0. (c) Temperature dependence of three peaks voltage for $d_S=63\,$nm and $d_N=34.5\,$nm, compared to the BCS gap dependence (lines are traced for V(T)=6.1$\times \Delta (T)$, 6.9$\times \Delta (T)$ and 7.3$\times \Delta (T)$). (d) Differential resistance curve for $d_S=57\,$nm and $d_N=23\,$nm as a function of the magnetic field at $T$=100 mK. }
\label{pics}
\end{figure}

\emph{Coherence and superconducting interferences--} We now describe the transport measurements realised on the bilayers through two contacts deposited on top of the superconducting layer. When increasing the current, we observe a switching from the superconducting to the resistive state at a critical current $I_c \sim 100-200 \, \mu$A, and a strong hysteresis when reducing the current, with a  retrapping current $I_r\sim 10-50 \,\mu$A. This hysteresis can be attributed to typical heating effects, that will be treated in more detail further in the paper.  \\
Fig. \ref{pics}-a shows the low current part of a typical differential resistance curve $dV/dI (I)$. Reproducible resistance peaks are observed in the resistive state of all the measured bilayers. We have studied the voltage, temperature and magnetic field dependence of these peaks to identify their origin.\\
Fig. \ref{pics}-b shows the voltage of the first 7 peaks for two bilayers with different $d_N$. 
For each $d_N$, the peaks are equally spaced in voltage (successive peaks having thus linearly increasing voltages), with $\Delta$V=3-17 $\mu$V in the temperature range studied 40-250 mK. When decreasing $d_N$, we observe an increase of $\Delta$V. For $d_{N,1}=26\,$nm and $d_{N,2}=22\,$nm we found respectively a periodicity $\Delta V_1=9.7\,\mu$V and $\Delta V_2=13.6\,\mu$V. This would suggest a dependence $\Delta V \propto 1/d_N^2$. Indeed, $(d_{N,1}/d_{N,2})^2 = 1.4 = \Delta V_2/\Delta V_1$. \\
Fig. \ref{pics}-c shows the temperature dependence of the peaks voltage, as compared to the temperature dependence of the BCS gap. 
The peaks disappear with temperature, and their dependence follows that of the BCS gap, demonstrating that the peaks are related to superconductivity.\\
A magnetic field rapidly suppresses the peaks, which cannot be observed for $H>H_c/3$ (Fig. \ref{pics}-d). They thus show a greater sensitivity to the magnetic field than the critical current, which is suppressed over the same magnetic field range by a factor 3.5 only.  \\
These periodic peaks, due to superconducting interferences, may be explained in the frame of the Rowell-McMillan effect. In a bilayer formed by a superconductor and a normal metal with a good interface, an electron in the normal metal is reflected at the S/N interface into a coherent hole through an Andreev reflection. The hole can be specularly reflected at the N/substrate interface, and undergoes a second Andreev reflection at the superconductor interface. The interference between the first electron, and the electron resulting from two Andreev reflections and one specular reflection, can induce a series of equidistant peaks in the bilayer resistance. 
The expected peaks periodicity in a ballistic system is $\Delta V=\hbar v_{F,N}/4 d_N e$, as the charges cross 4 times the thickness of the normal layer in a ballistic time $\tau_b=4 d_N/v_{F,N}$. As in our case the normal layer is in the dirty limit, the diffusive time is given by $\tau_D =(4 d_N)^2/D$, leading to $\Delta V \propto 1/d_N^2$. 
The experimental values found are also compatible with this interpretation: for $d_N=22\,$nm we find $\Delta V=\hbar D/(4 d_N)^2 e \sim 16 \mu$V, for a diffusion coefficient $D \sim 2\,cm^2 s$ calculated from $D=v_{F,N}\, l_e/3$ with $v_{F,N}=b\times v_{F,S}=1.2\times 10^5$m/s and $l_e=5\,$nm. This estimate is in rough agreement with the observed periodicity of 13.6 $\mu V$.\\
We expect that these oscillations, rarely observed in metallic systems, are enhanced in our devices thanks to the sharp epitaxial interface between the Si substrate and the Si:B crystalline N layer where the specular reflection takes place, illustrating again the quality of the realised structures. It is also worth noting that the peaks are observed up to the seventh order of interference; this implies that the phase coherence length is not strongly limited by inelastic diffusion phenomena, and extends at least to a few hundreds of nanometers.\\

\begin{figure}[t!]
\includegraphics[width=\columnwidth]{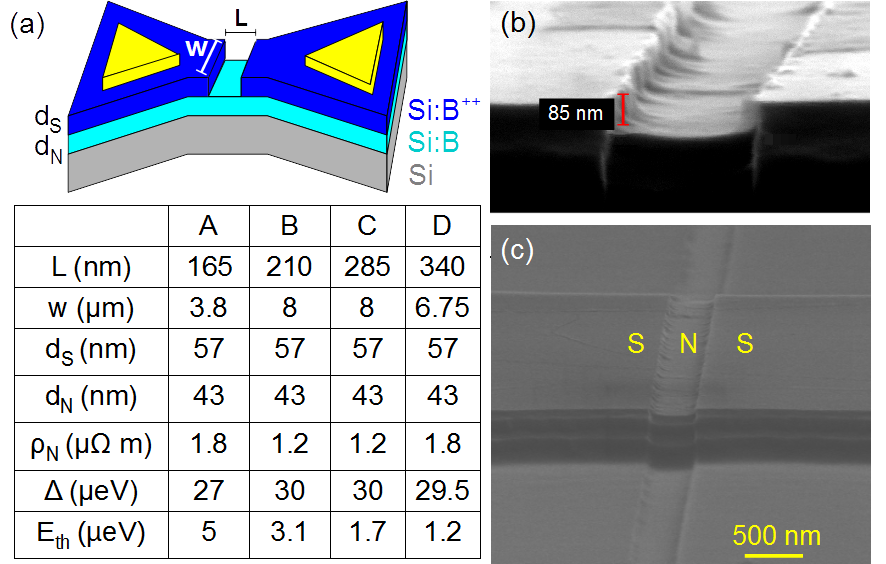}
\caption{(a) Sketch of the SNS junctions geometry realised from a S/N (Si:B$^{++}$/Si:B) bilayer; (b)-(c) SEM images of the line etched in the upper superconducting layer and partially in the normal metal layer. (Table) Characteristics of the junctions shown in this paper: normal wire length $L$ and width $w$, superconductor and normal metal thickness $d_N$ and $d_S$, normal metal resistivity $\rho$, gap of the superconducting bilayer $\Delta=1.78\,k_B T_c$, and Thouless energy $E_{Th}=\hbar D/L^2$, calculated from $L$ and the fitting value of the diffusion coefficient $D$. }
\label{SNSmeb}
\end{figure}

\emph{Josephson junctions --} Josephson junctions were realised from SN bilayers with $d_S=57\,$nm, $d_N=58\,$nm, $T_c=200\,$mK and a normal layer concentration of $5.2\times 10^{20}\,$cm$^{-3}$. 
The device contacts and the wire width $w: 3.8 - 8 \mu$m were defined by Reactive Ion Etching (RIE). Then, a thin line was drawn across the wire and the superconducting upper layer was etched by RIE, along with the first 15 nm of the normal layer, reducing $d_N$ to $43\,$nm (Fig. \ref{SNSmeb}). This insures the complete suppression of the superconductor in the weak link, and that there are no superconducting bridges left. The normal layer bridge length was set in the range $L:165 - 340\,$ nm (junctions A,B,C,D), giving a Thouless energy $E_{Th}= 1.2-5 \,\mu$eV$<\Delta \sim 30\,\mu$eV (Table, Fig. \ref{SNSmeb}). As $\Delta \sim 10\, E_{Th}$, our junctions are in the intermediate regime between the long and the short junction limit \cite{Dubos}.

\begin{figure}[t!]
\includegraphics[width=\columnwidth]{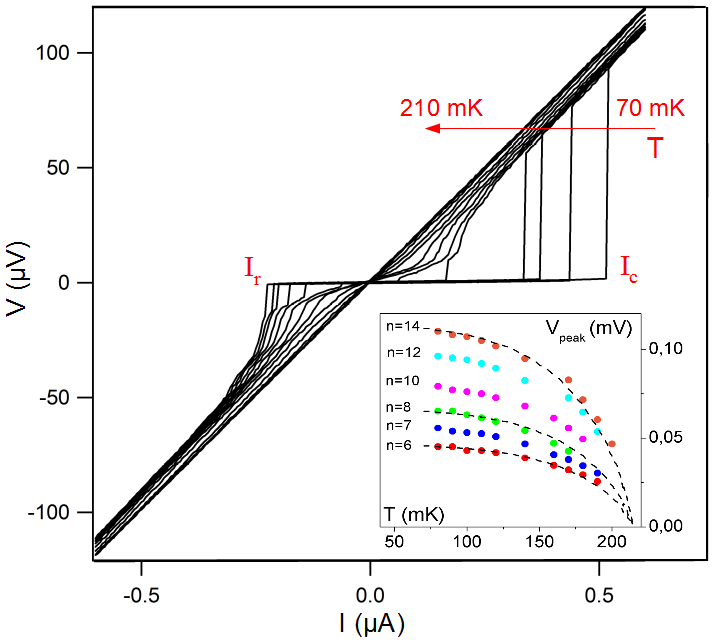}
\caption{(a) $V(I)$ curves for junction A in the temperature range 70-210 K for increasing current. (Inset): Temperature dependence of the peaks voltage, compared to the BCS gap dependence (dotted lines for 1.4$\times \Delta (T)$, 2$\times \Delta (T)$ and 3.4$\times \Delta (T)$).  }
\label{VI}
\end{figure}

\begin{figure}[t!]
\includegraphics[width=\columnwidth]{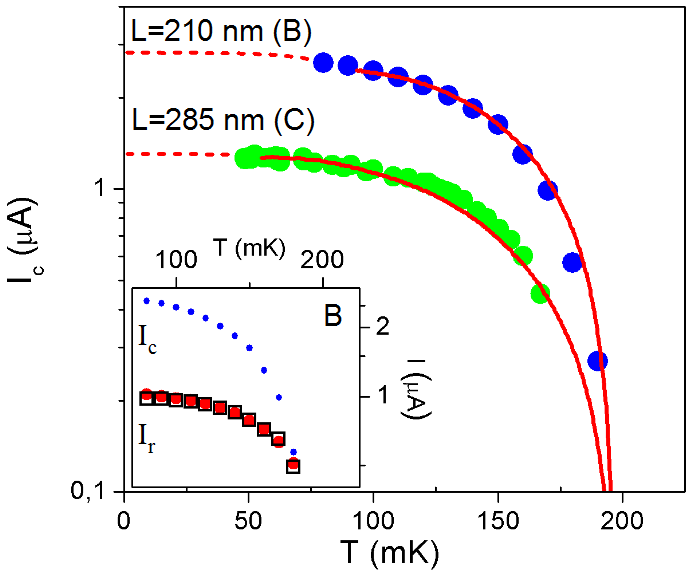}
\caption{$I_c (T)$ for junctions B and C. Lines: fits from eq. \ref{eqIcT}. Dotted lines: zero temperature limit. Inset: $I_c (T)$ and $I_r (T)$ for junction B. $I_r(T)$ fit (squares) assuming a heating-generated hysteresis.}
\label{IcT}
\end{figure}

The junctions $V(I)$ curves (Fig. \ref{VI}) show standard characteristics: a sharp jump when the junction switches to the resistive state at the critical current $I_c \sim 0.5-3\,\mu$A, and an hysteresis cycle when, reducing the bias current, the junction switches back to the superconducting state at the retrapping current $I_r <$1$\mu$A. As in the bilayers, multiple plateaus appear in the normal region of the V-I curves (as equidistant peaks appear in the dV/dI curve), whose origin can be traced back to the bilayers. This is supported by the temperature dependence of the plateaus (Fig. \ref{VI}-inset), which follows, as in the case of the bilayers, the dependence of $\Delta (T)$.\\
Fig. \ref{IcT} (a) displays the temperature dependence of the critical current in two junctions (B and C) of different length $L=210\,$nm and $L=285\,$nm (Fig. \ref{IcT}-a).
The critical current dependence on temperature has been fitted with the de Gennes expression for a dirty doped semiconductor weak link. This expression is valid in the linear limit for small induced pair correlations ($T\gtrsim T_c/2 \sim 100\,$mK) \cite{Seto}:
\begin{equation}
I_c (T) \propto \frac{1}{\xi_N}\,\left(\frac{\Delta (T)}{cosh(L/2\xi_N)}\right)^2
\label{eqIcT}
\end{equation}
The coherence length in the normal wire, $\xi_N$, is dominated by the thermal coherence length, and was calculated following:
\begin{equation}
\xi_N=\sqrt{\frac{\hbar^3\,\mu}{6 \pi k_B T e m^{*}}}\,(3 \pi^2 n_N)^{1/3}
\end{equation}
with $m^{*}=0.34\, m_e$, $n_N=5.2\times 10^{20}$cm$^{-3}$ and $\mu=55\,$cm$^2$/V s, in reasonable agreement with the mobility estimated from the measured resistivity at low temperature $\mu=1/e n_N \rho \sim 60\,$cm$^2$/V s. The calculated coherence length in the normal weak link at $T_c=197\,$mK is $\xi_N (T_c) \sim 115 \,$nm, larger than expected in metallic Si:B, as already observed \cite{serfaty}, due probably to the fact that Si:B becomes itself superconducting at $T_c=0.07\,$K.\\
The $I_c (T)$ fitting allows to extract the coherence length in the normal wire; furthermore, we can extract the Thouless energy from the low temperature value of $I_c$.
Indeed, while in a short junction the critical current is expected to saturate at low temperature at $e\,\,I_c (0) = 2.08 \, \Delta (0)$ \cite{Dubos}, we are here in an intermediate regime between the short and the long junction limit. In this case, $e\,R_N\,I_c (0) = a \, \Delta (0)$, with $a$ decreasing monotonically and smoothly when the ratio $E_{Th}/\Delta$ decreases towards the opposite limit, the long junction limit, where $\Delta>> E_{Th}$ and $e\,R_N\,I_c (0) = 10.82 \, E_{Th}$ \cite{Dubos}. We can thus extract the value of $E_{Th}$ from $a$. For junction B, we find $a=0.64$ and $E_{Th}/\Delta=0.105$ while for junction C, we find $a=0.42$ and $E_{Th}/\Delta=0.057$. Assuming the same diffusion coefficient in both junctions, we confirm that $E_{Th,B}/E_{Th,C}=1.84 =(L_C/L_B)^2$. Moreover, the diffusion coefficient deduced from $E_{Th}$, $D=2.2\,cm^2/s$, is in good agreement with our previous estimate ($D=2\,cm^2/s$). Such confirmation is important, as parameters such as $l_e$ or $v_{F}$ are not well known for the peculiar material that is ultra-doped silicon, but are of a fundamental importance to estimate the performances of superconducting silicon-based devices.  \\

The inset of Fig. \ref{IcT} shows the temperature dependence of the  retrapping current $I_r$, compared to $I_c(T)$ in junction B. At low temperature, an hysteresis can be observed, as the switching current from the normal to the superconducting state, $I_r$, is smaller than the critical current $I_c$; at $T>180\,$mK, the hysteresis disappears as $I_c=I_r$.
SNS junctions often present similar hysteresis at low temperature, due to the heating of the junction by the Joule power generated in the resistive state. At low temperature, the phonon cooling power is limited, and the injected heat cannot be instantly dissipated. The electronic temperature $T_{el}$ is then larger than the phonon temperature $T_{ph}$, and the junction can switch back in the superconducting state only at $I_r=I_c(T_{el})<I_c(T_{ph})$. Our observations confirm such heating effect: the hysteresis disappears at higher temperatures, and we can fit the retrapping current evolution by assuming a phonon cooling power $P=\Sigma\,V\,(T_{el}^5-T_{ph}^5)$ with $\Sigma=5.2\times 10^{7} W/K^{5}\, m^3$. This value is in good agreement with the phonon cooling power measured for n-Si films of similar strong doping \cite{giazotto}.\\

\emph{Conclusions --} 
We have realised all-silicon superconducting/ normal metal bilayers of variable thickness and doping, that we have successfully modelled in the frame of diffusive metallic superconductors. We demonstrated a transparent S/N interface, an asset of the epitaxial sharp interface realised, with an interface resistance lower by an order of magnitude than metal/doped Si contact resistances. In this well controlled, crystalline system, we also observed differential resistance peaks, equally spaced in voltage. These superconductivity-induced interferences, visible thanks to a long phase coherence length, are consistent with the Rowell-McMillan effect, but further investigation is necessary to confirm this hypothesis. Finally, we exploited the low resistance S/N interfaces to realise all-silicon lateral SNS Josephson junctions from the SN bilayers. We observed a full proximity effect with a critical current of a few $\mu$A. The transport features, such as the I-V characteristic or the temperature dependence of the critical and retrapping current, could be well understood and modelled. Furthermore, this work lead to the estimation of important parameters in ultra-doped superconducting Si, such as the Fermi velocity, the coherence length, or the electron-phonon coupling constant, extremely useful to estimate the performances of future superconducting silicon-based devices.\\

\emph{Supplementary materials --} We examined the behaviour of both bilayers and junctions in a perpendicular magnetic field.

\begin{figure}[t!]
\includegraphics[width=0.9\columnwidth]{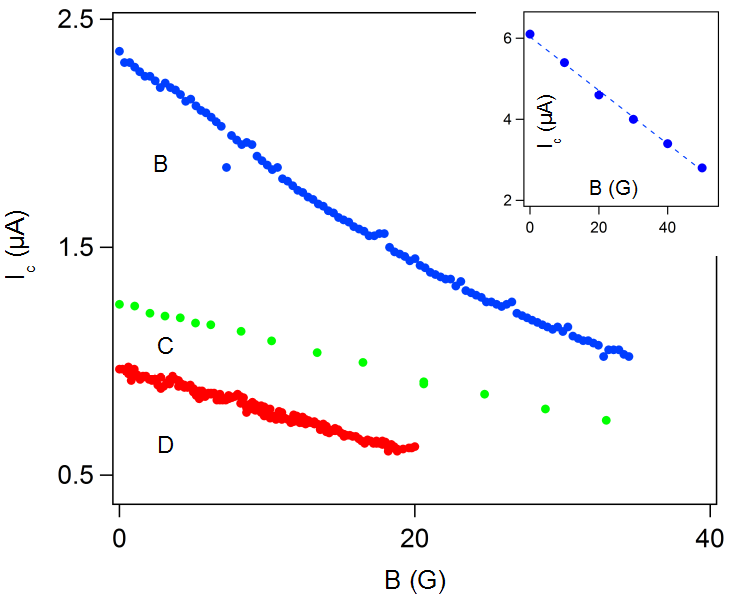}
\caption{$I_c$ vs B for junctions B ($L\times w$=1.68$\mu$m$^2$), C ($L\times w$=2.28$\mu$m$^2$) and D ($L\times w$=2.3$\mu$m$^2$). Inset: $I_c$ vs $B$ for a wide section of a SN bilayer ($60\,\mu m$ long and $160\,\mu m$ wide), with $d_S=57 \,$nm and $d_N=58\,$nm. The linear behaviour follows $I_c/B=-0.066 \mu A/G$. }
\label{IcH}
\end{figure}

Fig. \ref{IcH} shows the $I_c$ vs $H$ dependence of three junctions of different length. All show an overall linear dependence as the critical current decreases with the field in a range of a few tens of gauss. 
A wide section of a SN bilayer ($60\,\mu m$ long and $160\,\mu m$ wide) is also shown for comparison, and presents a linear decrease of the critical current with the magnetic field over a comparable range. The superconductivity in the bilayer, and thus in the junctions' contact pads, is quickly suppressed by the magnetic field, reducing at the same time the junctions' critical current. This envelope dominates the junctions with the weakest $I_c$, while a rounded shape can be perceived in junction B. To explain the absence of the minima of the expected Fraunhofer pattern, that should be still visible despite the envelope, we estimated the importance of the self field effects. Indeed, the width of the junctions studied (3.8-8 $\mu$m) is of the same order of magnitude of the Josephson length \cite{Barone}. 
\begin{equation}
\lambda_J = \sqrt{\frac{\hbar}{2 e \mu_0 (L+2\lambda_L) J_c}}
\end{equation}
where $\lambda_L \sim 120 \,nm$ is the London penetration length for the superconducting Si:B layer calculated from the previous estimations.
For the three junctions shown we have, respectively, $\lambda_J= 7.9\,\mu m \sim w=8\,\mu m$ (junction B); $\lambda_J= 10\,\mu m \gtrsim w=8\,\mu m$ (junction C); $\lambda_J= 10\,\mu m > w=6.75\,\mu m$ (junction D).
The self field tends to induce a linear dependence of the $I_c(H)$ curve, but the amplitude of the effect is not sufficient to cancel completely the Fraunhofer minima \cite{Barone}. We expect that a strong inhomogeneity of the current lines distribution may further contribute to the monotonic suppression observed.\\

We are grateful to J.R. Coudevylle for help with the fabrication, and to H. Cercellier, M. V. Feigel'man, S. Guéron and T. Klein for fruitful discussions.
This work was supported by the R\'eseau RENATECH, the FCS Triangle de la Physique through the project 'DownTo40', the ANR 'Nanoquartet', and the CEA program 'Transverse Nanoscience'.


\end{document}